\documentclass{PoS}
\usepackage{graphics}

\PoS{PoS(LAT2005)179}

\title{Casimir Scaling of domain wall tensions in the deconfined phase of D=3+1
SU(N) gauge theories. }

\ShortTitle{Casimir Scaling of domain wall tensions. }

\author{\speaker{Francis Bursa}\\
        University of Oxford\\
        E-mail: \email{bursa@thphys.ox.ac.uk}}
            
\author{Michael Teper\\
        University of Oxford\\
        E-mail: \email{teper@thphys.ox.ac.uk}}

\abstract{We perform lattice calculations of the spatial 't Hooft
$k$-string tensions, $\tilde\sigma_k$, in the deconfined phase of SU($N$)
gauge theories for $N=2,3,4,6$. These equal
(up to a factor of $T$) the surface tensions of the domain walls
between the corresponding (Euclidean) deconfined phases.
For $T\gg T_c$ our results match on to the known perturbative result,
which exhibits Casimir Scaling, $\tilde\sigma_k \propto k(N-k)$.
At lower $T$ the coupling becomes stronger and, not surprisingly, our
calculations show large deviations from the perturbative $T$-dependence.
Despite this we find that the behaviour
$\partial \tilde\sigma_k/\partial T \propto k(N-k)$ persists very accurately down to
temperatures very close to $T_c$. Thus the Casimir Scaling of the
't Hooft tension appears to be a `universal' feature that
is more general than its origin in the low order high-$T$
perturbative calculation. We observe the `wetting' of
these $k$-walls at $T\simeq T_c$. 
Our calculations show that as
$T\rightarrow T_c$ the magnitude of $\tilde\sigma_k(T)$ decreases rapidly.
}         

\FullConference{XXIIIrd International Symposium on Lattice Field Theory\\
                 25-30 July 2005\\
                 Trinity College, Dublin, Ireland}

\begin{document}

\section{Introduction}
In the Euclidean formulation of SU($N$) gauge theories at finite
$T$, deconfinement is associated with the 
spontaneous breaking of a $Z_N$ centre symmetry. 
The obvious order parameter is 
the trace of the Polyakov loop, $l_p$. In the
deconfined phase the effective potential for the Polyakov loop trace
averaged over the volume, $\bar{l}_p$, has its minimum at 
$\bar{l}_p {\not =} 0$, so the symmetry is spontaneously
broken and there are $N$ possible deconfined phases. If 
$\bar{l}_p \propto z_k=e^{2\pi i\frac{k}{N}}$ we label the phase by $k$. When
two of these phases, $k_1$ and $k_2$ say, co-exist they 
will be separated by a domain wall whose surface tension 
$\sigma_W^k$ will depend on $k=k_1-k_2$, as well as on $N$ and $T$.  
The tension of these domain walls is equal, up to a factor of $T$,
to the spatial 't Hooft string tension, which is in principle related 
to confinement.

At high $T$ one can calculate $\sigma_W^k$ in
perturbation theory. To two loops one finds
\cite{CKA1,CKA2}:
\begin{equation}
\sigma_W^k
=
k(N-k)
\frac{4\pi^2}{3\sqrt{3}}
\frac{T^3}{\sqrt{g^2(T) N}}
\{1 - \tilde{c}_2 g^2(T)N\}
\label{eqn_sigkPT}
\end{equation}
where $\tilde{c}_2 \simeq  0.09$. 

The factor $k(N-k)$ is the $k$ dependence
of the Casimir, $Tr_{\cal{R}}T^aT^a$, where 
$\cal{R}$ is the totally antisymmetric representation of
a product of $k$ fundamentals of SU($N$). It is the factor 
one obtains when calculating the Coulomb interaction
between sources in such a representation. In $D$=1+1 SU($N$) gauge
theories, the tension of the confining $k$-string that connects
such sources has precisely this dependence. There are 
old speculations
\cite{CS}
that this `Casimir Scaling' holds in $D$=3+1 and numerical
calculations show that it is a good (but not exact) 
approximation 
\cite{oxsigk}.
A question we will try to answer in this paper is whether
the Casimir Scaling in eqn(\ref{eqn_sigkPT}) survives 
at very much lower $T$.
This question connects to the role of Casimir Scaling in confinement,
since these domain walls are closely related to the centre vortices
that provide a possible mechanism for confinement\cite{tHooft,jgrev}.

%

\section{Preliminaries}
We discretise Euclidean space-time to a periodic hypercubic
lattice.
The $\mu=0$ direction is the temperature direction
and the domain wall spans the $L_1\times L_2$ torus.

Our order parameter will be based on the Polyakov loop $l_p$.
Above $T_c$ there are $N$ degenerate phases in which
$\langle l_p \rangle = z_k c(\beta)$ where $c(\beta)$ is a 
real-valued renormalisation factor. They can
co-exist at any $T\geq T_c$ and will be separated 
by domain walls. If $l_p$ in the two phases differs
by a factor $z_k$ we refer to the domain wall as a $k$-wall.

To study a $k$-wall we use a `twisted' plaquette action
to enforce the presence of a single domain
wall. The twisted action is defined by
\begin{equation}
S_{k}
=
\sum_p \Bigl(1-\frac{1}{N} \mathrm{ReTr} \{z(p) U_p\}\Bigr),
\label{eqn_Stw}
\end{equation}
where  $z(p)=1$ for all plaquettes except
\begin{equation}
z(p=\{\mu\nu,x_\mu\})
= z_k =
e^{2\pi i\frac{k}{N}}
 \  \  \quad \mu\nu=03;  \  x_0={x_0}^\prime,
 \  x_3={x_3}^\prime, \  x_{1,2}=1,...,L_{1,2}.
\label{eqn_Stw2}
\end{equation}

That is to say, the plaquettes in the entire (0,3)-plane at
$x_0={x_0}^\prime$ and  $x_3={x_3}^\prime$
are multiplied by $z_k\in Z_N$.
The Polyakov loops on either side of the plane will differ by a factor of $z_k$. 
Periodicity in $x_3$ then demands that at some $x_3$ the
Polyakov loop must suffer a compensating factor of
${z_k}^{\dagger}$ -- a domain wall. Thus we
ensure that each configuration possesses
at least one $k$-wall.

We calculate the average action with and without a
$k$-twist as defined above. The difference is
\begin{equation}
\Delta S_k 
\equiv 
\langle S_k \rangle - \langle S_0 \rangle
=
\frac{\partial \ln Z_0}{\partial\beta}
-
\frac{\partial \ln Z_k}{\partial\beta}
=
\frac{\partial}{\partial\beta}\frac{F_k-F_0}{T}
=
\frac{\partial}{\partial\beta}\frac{\sigma_W^k A}{T}
\label{eqn_Swall}
\end{equation}
where $\sigma_W^k$ is the surface tension 
of the domain wall and $A=a^2L_1L_2$ is its area.

We see from eqn(\ref{eqn_sigkPT}) that we expect 
\begin{equation}
\Delta S_k 
=
\frac{\partial}{\partial\beta}\frac{\sigma_W^k A}{T}
\stackrel{T\to \infty}{=}
 \alpha(L_0) \frac{k(N-k)}{\sqrt{N}} 
\frac{4\pi^2}{3\sqrt{3}}
\frac{L_1L_2}{L^2_0}
\frac{\partial}{\partial\beta}
\frac{1}{g(T)}
\label{eqn_SkPT1}
\end{equation}
when leading order perturbation 
theory is accurate. The factor of  $\alpha(L_0)$ 
contains the $O(a^2T^2) = O(1/L^2_0)$ lattice correction, obtained 
by a numerical evaluation of the expression 
given in \cite{DWd3}.




\section{Results}
\subsection{Surface tension}

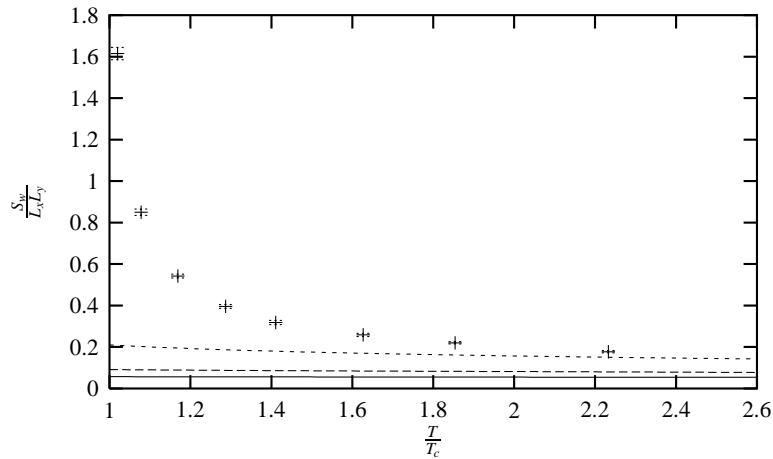
\begin	{figure}[p]
\begin	{center}
\leavevmode
\scalebox{0.8}{
\begingroup%
  \makeatletter%
  \newcommand{\GNUPLOTspecial}{%
    \@sanitize\catcode`\%=14\relax\special}%
  \setlength{\unitlength}{0.1bp}%
{\GNUPLOTspecial{!
/gnudict 256 dict def
gnudict begin
/Color false def
/Solid false def
/gnulinewidth 5.000 def
/userlinewidth gnulinewidth def
/vshift -33 def
/dl {10 mul} def
/hpt_ 31.5 def
/vpt_ 31.5 def
/hpt hpt_ def
/vpt vpt_ def
/M {moveto} bind def
/L {lineto} bind def
/R {rmoveto} bind def
/V {rlineto} bind def
/vpt2 vpt 2 mul def
/hpt2 hpt 2 mul def
/Lshow { currentpoint stroke M
  0 vshift R show } def
/Rshow { currentpoint stroke M
  dup stringwidth pop neg vshift R show } def
/Cshow { currentpoint stroke M
  dup stringwidth pop -2 div vshift R show } def
/UP { dup vpt_ mul /vpt exch def hpt_ mul /hpt exch def
  /hpt2 hpt 2 mul def /vpt2 vpt 2 mul def } def
/DL { Color {setrgbcolor Solid {pop []} if 0 setdash }
 {pop pop pop Solid {pop []} if 0 setdash} ifelse } def
/BL { stroke userlinewidth 2 mul setlinewidth } def
/AL { stroke userlinewidth 2 div setlinewidth } def
/UL { dup gnulinewidth mul /userlinewidth exch def
      10 mul /udl exch def } def
/PL { stroke userlinewidth setlinewidth } def
/LTb { BL [] 0 0 0 DL } def
/LTa { AL [1 udl mul 2 udl mul] 0 setdash 0 0 0 setrgbcolor } def
/LT0 { PL [] 1 0 0 DL } def
/LT1 { PL [4 dl 2 dl] 0 1 0 DL } def
/LT2 { PL [2 dl 3 dl] 0 0 1 DL } def
/LT3 { PL [1 dl 1.5 dl] 1 0 1 DL } def
/LT4 { PL [5 dl 2 dl 1 dl 2 dl] 0 1 1 DL } def
/LT5 { PL [4 dl 3 dl 1 dl 3 dl] 1 1 0 DL } def
/LT6 { PL [2 dl 2 dl 2 dl 4 dl] 0 0 0 DL } def
/LT7 { PL [2 dl 2 dl 2 dl 2 dl 2 dl 4 dl] 1 0.3 0 DL } def
/LT8 { PL [2 dl 2 dl 2 dl 2 dl 2 dl 2 dl 2 dl 4 dl] 0.5 0.5 0.5 DL } def
/Pnt { stroke [] 0 setdash
   gsave 1 setlinecap M 0 0 V stroke grestore } def
/Dia { stroke [] 0 setdash 2 copy vpt add M
  hpt neg vpt neg V hpt vpt neg V
  hpt vpt V hpt neg vpt V closepath stroke
  Pnt } def
/Pls { stroke [] 0 setdash vpt sub M 0 vpt2 V
  currentpoint stroke M
  hpt neg vpt neg R hpt2 0 V stroke
  } def
/Box { stroke [] 0 setdash 2 copy exch hpt sub exch vpt add M
  0 vpt2 neg V hpt2 0 V 0 vpt2 V
  hpt2 neg 0 V closepath stroke
  Pnt } def
/Crs { stroke [] 0 setdash exch hpt sub exch vpt add M
  hpt2 vpt2 neg V currentpoint stroke M
  hpt2 neg 0 R hpt2 vpt2 V stroke } def
/TriU { stroke [] 0 setdash 2 copy vpt 1.12 mul add M
  hpt neg vpt -1.62 mul V
  hpt 2 mul 0 V
  hpt neg vpt 1.62 mul V closepath stroke
  Pnt  } def
/Star { 2 copy Pls Crs } def
/BoxF { stroke [] 0 setdash exch hpt sub exch vpt add M
  0 vpt2 neg V  hpt2 0 V  0 vpt2 V
  hpt2 neg 0 V  closepath fill } def
/TriUF { stroke [] 0 setdash vpt 1.12 mul add M
  hpt neg vpt -1.62 mul V
  hpt 2 mul 0 V
  hpt neg vpt 1.62 mul V closepath fill } def
/TriD { stroke [] 0 setdash 2 copy vpt 1.12 mul sub M
  hpt neg vpt 1.62 mul V
  hpt 2 mul 0 V
  hpt neg vpt -1.62 mul V closepath stroke
  Pnt  } def
/TriDF { stroke [] 0 setdash vpt 1.12 mul sub M
  hpt neg vpt 1.62 mul V
  hpt 2 mul 0 V
  hpt neg vpt -1.62 mul V closepath fill} def
/DiaF { stroke [] 0 setdash vpt add M
  hpt neg vpt neg V hpt vpt neg V
  hpt vpt V hpt neg vpt V closepath fill } def
/Pent { stroke [] 0 setdash 2 copy gsave
  translate 0 hpt M 4 {72 rotate 0 hpt L} repeat
  closepath stroke grestore Pnt } def
/PentF { stroke [] 0 setdash gsave
  translate 0 hpt M 4 {72 rotate 0 hpt L} repeat
  closepath fill grestore } def
/Circle { stroke [] 0 setdash 2 copy
  hpt 0 360 arc stroke Pnt } def
/CircleF { stroke [] 0 setdash hpt 0 360 arc fill } def
/C0 { BL [] 0 setdash 2 copy moveto vpt 90 450  arc } bind def
/C1 { BL [] 0 setdash 2 copy        moveto
       2 copy  vpt 0 90 arc closepath fill
               vpt 0 360 arc closepath } bind def
/C2 { BL [] 0 setdash 2 copy moveto
       2 copy  vpt 90 180 arc closepath fill
               vpt 0 360 arc closepath } bind def
/C3 { BL [] 0 setdash 2 copy moveto
       2 copy  vpt 0 180 arc closepath fill
               vpt 0 360 arc closepath } bind def
/C4 { BL [] 0 setdash 2 copy moveto
       2 copy  vpt 180 270 arc closepath fill
               vpt 0 360 arc closepath } bind def
/C5 { BL [] 0 setdash 2 copy moveto
       2 copy  vpt 0 90 arc
       2 copy moveto
       2 copy  vpt 180 270 arc closepath fill
               vpt 0 360 arc } bind def
/C6 { BL [] 0 setdash 2 copy moveto
      2 copy  vpt 90 270 arc closepath fill
              vpt 0 360 arc closepath } bind def
/C7 { BL [] 0 setdash 2 copy moveto
      2 copy  vpt 0 270 arc closepath fill
              vpt 0 360 arc closepath } bind def
/C8 { BL [] 0 setdash 2 copy moveto
      2 copy vpt 270 360 arc closepath fill
              vpt 0 360 arc closepath } bind def
/C9 { BL [] 0 setdash 2 copy moveto
      2 copy  vpt 270 450 arc closepath fill
              vpt 0 360 arc closepath } bind def
/C10 { BL [] 0 setdash 2 copy 2 copy moveto vpt 270 360 arc closepath fill
       2 copy moveto
       2 copy vpt 90 180 arc closepath fill
               vpt 0 360 arc closepath } bind def
/C11 { BL [] 0 setdash 2 copy moveto
       2 copy  vpt 0 180 arc closepath fill
       2 copy moveto
       2 copy  vpt 270 360 arc closepath fill
               vpt 0 360 arc closepath } bind def
/C12 { BL [] 0 setdash 2 copy moveto
       2 copy  vpt 180 360 arc closepath fill
               vpt 0 360 arc closepath } bind def
/C13 { BL [] 0 setdash  2 copy moveto
       2 copy  vpt 0 90 arc closepath fill
       2 copy moveto
       2 copy  vpt 180 360 arc closepath fill
               vpt 0 360 arc closepath } bind def
/C14 { BL [] 0 setdash 2 copy moveto
       2 copy  vpt 90 360 arc closepath fill
               vpt 0 360 arc } bind def
/C15 { BL [] 0 setdash 2 copy vpt 0 360 arc closepath fill
               vpt 0 360 arc closepath } bind def
/Rec   { newpath 4 2 roll moveto 1 index 0 rlineto 0 exch rlineto
       neg 0 rlineto closepath } bind def
/Square { dup Rec } bind def
/Bsquare { vpt sub exch vpt sub exch vpt2 Square } bind def
/S0 { BL [] 0 setdash 2 copy moveto 0 vpt rlineto BL Bsquare } bind def
/S1 { BL [] 0 setdash 2 copy vpt Square fill Bsquare } bind def
/S2 { BL [] 0 setdash 2 copy exch vpt sub exch vpt Square fill Bsquare } bind def
/S3 { BL [] 0 setdash 2 copy exch vpt sub exch vpt2 vpt Rec fill Bsquare } bind def
/S4 { BL [] 0 setdash 2 copy exch vpt sub exch vpt sub vpt Square fill Bsquare } bind def
/S5 { BL [] 0 setdash 2 copy 2 copy vpt Square fill
       exch vpt sub exch vpt sub vpt Square fill Bsquare } bind def
/S6 { BL [] 0 setdash 2 copy exch vpt sub exch vpt sub vpt vpt2 Rec fill Bsquare } bind def
/S7 { BL [] 0 setdash 2 copy exch vpt sub exch vpt sub vpt vpt2 Rec fill
       2 copy vpt Square fill
       Bsquare } bind def
/S8 { BL [] 0 setdash 2 copy vpt sub vpt Square fill Bsquare } bind def
/S9 { BL [] 0 setdash 2 copy vpt sub vpt vpt2 Rec fill Bsquare } bind def
/S10 { BL [] 0 setdash 2 copy vpt sub vpt Square fill 2 copy exch vpt sub exch vpt Square fill
       Bsquare } bind def
/S11 { BL [] 0 setdash 2 copy vpt sub vpt Square fill 2 copy exch vpt sub exch vpt2 vpt Rec fill
       Bsquare } bind def
/S12 { BL [] 0 setdash 2 copy exch vpt sub exch vpt sub vpt2 vpt Rec fill Bsquare } bind def
/S13 { BL [] 0 setdash 2 copy exch vpt sub exch vpt sub vpt2 vpt Rec fill
       2 copy vpt Square fill Bsquare } bind def
/S14 { BL [] 0 setdash 2 copy exch vpt sub exch vpt sub vpt2 vpt Rec fill
       2 copy exch vpt sub exch vpt Square fill Bsquare } bind def
/S15 { BL [] 0 setdash 2 copy Bsquare fill Bsquare } bind def
/D0 { gsave translate 45 rotate 0 0 S0 stroke grestore } bind def
/D1 { gsave translate 45 rotate 0 0 S1 stroke grestore } bind def
/D2 { gsave translate 45 rotate 0 0 S2 stroke grestore } bind def
/D3 { gsave translate 45 rotate 0 0 S3 stroke grestore } bind def
/D4 { gsave translate 45 rotate 0 0 S4 stroke grestore } bind def
/D5 { gsave translate 45 rotate 0 0 S5 stroke grestore } bind def
/D6 { gsave translate 45 rotate 0 0 S6 stroke grestore } bind def
/D7 { gsave translate 45 rotate 0 0 S7 stroke grestore } bind def
/D8 { gsave translate 45 rotate 0 0 S8 stroke grestore } bind def
/D9 { gsave translate 45 rotate 0 0 S9 stroke grestore } bind def
/D10 { gsave translate 45 rotate 0 0 S10 stroke grestore } bind def
/D11 { gsave translate 45 rotate 0 0 S11 stroke grestore } bind def
/D12 { gsave translate 45 rotate 0 0 S12 stroke grestore } bind def
/D13 { gsave translate 45 rotate 0 0 S13 stroke grestore } bind def
/D14 { gsave translate 45 rotate 0 0 S14 stroke grestore } bind def
/D15 { gsave translate 45 rotate 0 0 S15 stroke grestore } bind def
/DiaE { stroke [] 0 setdash vpt add M
  hpt neg vpt neg V hpt vpt neg V
  hpt vpt V hpt neg vpt V closepath stroke } def
/BoxE { stroke [] 0 setdash exch hpt sub exch vpt add M
  0 vpt2 neg V hpt2 0 V 0 vpt2 V
  hpt2 neg 0 V closepath stroke } def
/TriUE { stroke [] 0 setdash vpt 1.12 mul add M
  hpt neg vpt -1.62 mul V
  hpt 2 mul 0 V
  hpt neg vpt 1.62 mul V closepath stroke } def
/TriDE { stroke [] 0 setdash vpt 1.12 mul sub M
  hpt neg vpt 1.62 mul V
  hpt 2 mul 0 V
  hpt neg vpt -1.62 mul V closepath stroke } def
/PentE { stroke [] 0 setdash gsave
  translate 0 hpt M 4 {72 rotate 0 hpt L} repeat
  closepath stroke grestore } def
/CircE { stroke [] 0 setdash 
  hpt 0 360 arc stroke } def
/Opaque { gsave closepath 1 setgray fill grestore 0 setgray closepath } def
/DiaW { stroke [] 0 setdash vpt add M
  hpt neg vpt neg V hpt vpt neg V
  hpt vpt V hpt neg vpt V Opaque stroke } def
/BoxW { stroke [] 0 setdash exch hpt sub exch vpt add M
  0 vpt2 neg V hpt2 0 V 0 vpt2 V
  hpt2 neg 0 V Opaque stroke } def
/TriUW { stroke [] 0 setdash vpt 1.12 mul add M
  hpt neg vpt -1.62 mul V
  hpt 2 mul 0 V
  hpt neg vpt 1.62 mul V Opaque stroke } def
/TriDW { stroke [] 0 setdash vpt 1.12 mul sub M
  hpt neg vpt 1.62 mul V
  hpt 2 mul 0 V
  hpt neg vpt -1.62 mul V Opaque stroke } def
/PentW { stroke [] 0 setdash gsave
  translate 0 hpt M 4 {72 rotate 0 hpt L} repeat
  Opaque stroke grestore } def
/CircW { stroke [] 0 setdash 
  hpt 0 360 arc Opaque stroke } def
/BoxFill { gsave Rec 1 setgray fill grestore } def
end
}}%
\begin{picture}(3600,2160)(0,0)%
{\GNUPLOTspecial{"
gnudict begin
gsave
0 0 translate
0.100 0.100 scale
0 setgray
newpath
1.000 UL
LTb
400 300 M
63 0 V
2987 0 R
-63 0 V
400 496 M
63 0 V
2987 0 R
-63 0 V
400 691 M
63 0 V
2987 0 R
-63 0 V
400 887 M
63 0 V
2987 0 R
-63 0 V
400 1082 M
63 0 V
2987 0 R
-63 0 V
400 1278 M
63 0 V
2987 0 R
-63 0 V
400 1473 M
63 0 V
2987 0 R
-63 0 V
400 1669 M
63 0 V
2987 0 R
-63 0 V
400 1864 M
63 0 V
2987 0 R
-63 0 V
400 2060 M
63 0 V
2987 0 R
-63 0 V
400 300 M
0 63 V
0 1697 R
0 -63 V
781 300 M
0 63 V
0 1697 R
0 -63 V
1163 300 M
0 63 V
0 1697 R
0 -63 V
1544 300 M
0 63 V
0 1697 R
0 -63 V
1925 300 M
0 63 V
0 1697 R
0 -63 V
2306 300 M
0 63 V
0 1697 R
0 -63 V
2687 300 M
0 63 V
0 1697 R
0 -63 V
3069 300 M
0 63 V
0 1697 R
0 -63 V
3450 300 M
0 63 V
0 1697 R
0 -63 V
1.000 UL
LTb
400 300 M
3050 0 V
0 1760 V
-3050 0 V
400 300 L
1.000 UL
LT0
400 356 M
10 0 V
28 0 V
27 0 V
28 0 V
28 -1 V
28 0 V
28 0 V
29 0 V
28 0 V
29 0 V
29 0 V
30 0 V
29 0 V
30 0 V
30 0 V
30 0 V
30 0 V
30 0 V
31 0 V
30 0 V
31 0 V
31 0 V
32 0 V
31 0 V
32 0 V
32 0 V
32 0 V
32 0 V
32 0 V
32 0 V
33 0 V
33 0 V
33 -1 V
33 0 V
33 0 V
33 0 V
33 0 V
34 0 V
34 0 V
34 0 V
34 0 V
34 0 V
34 0 V
34 0 V
35 0 V
35 0 V
34 0 V
35 0 V
35 0 V
35 0 V
35 0 V
36 0 V
35 0 V
35 0 V
36 0 V
36 0 V
35 0 V
36 0 V
36 0 V
36 0 V
36 0 V
36 -1 V
37 0 V
36 0 V
36 0 V
37 0 V
36 0 V
37 0 V
36 0 V
37 0 V
37 0 V
36 0 V
37 0 V
37 0 V
37 0 V
37 0 V
37 0 V
36 0 V
37 0 V
37 0 V
37 0 V
37 0 V
38 0 V
37 0 V
37 0 V
37 0 V
37 0 V
37 0 V
37 0 V
37 0 V
34 0 V
1.000 UL
LT1
400 389 M
31 0 V
31 0 V
30 -1 V
31 0 V
31 0 V
31 0 V
31 0 V
30 -1 V
31 0 V
31 0 V
31 0 V
31 0 V
31 -1 V
30 0 V
31 0 V
31 0 V
31 0 V
31 -1 V
30 0 V
31 0 V
31 0 V
31 0 V
31 0 V
30 -1 V
31 0 V
31 0 V
31 0 V
31 0 V
30 0 V
31 0 V
31 -1 V
31 0 V
31 0 V
30 0 V
31 0 V
31 0 V
31 0 V
31 -1 V
31 0 V
30 0 V
31 0 V
31 0 V
31 0 V
31 0 V
30 0 V
31 0 V
31 -1 V
31 0 V
31 0 V
30 0 V
31 0 V
31 0 V
31 0 V
31 0 V
30 0 V
31 -1 V
31 0 V
31 0 V
31 0 V
30 0 V
31 0 V
31 0 V
31 0 V
31 0 V
31 0 V
30 -1 V
31 0 V
31 0 V
31 0 V
31 0 V
30 0 V
31 0 V
31 0 V
31 0 V
31 -1 V
30 0 V
31 0 V
31 0 V
31 0 V
31 0 V
30 0 V
31 0 V
31 0 V
31 -1 V
31 0 V
30 0 V
31 0 V
31 0 V
31 0 V
31 0 V
31 0 V
30 -1 V
31 0 V
31 0 V
31 0 V
31 0 V
30 0 V
31 0 V
31 -1 V
1.000 UL
LT2
400 505 M
10 0 V
28 -2 V
27 -1 V
28 -1 V
28 -2 V
28 -1 V
28 -1 V
29 -2 V
28 -1 V
29 -1 V
29 -1 V
30 -1 V
29 -1 V
30 -2 V
30 -1 V
30 -1 V
30 -1 V
30 -1 V
31 -1 V
30 -1 V
31 -1 V
31 -1 V
32 -1 V
31 -1 V
32 0 V
32 -1 V
32 -1 V
32 -1 V
32 -1 V
32 -1 V
33 -1 V
33 0 V
33 -1 V
33 -1 V
33 -1 V
33 0 V
33 -1 V
34 -1 V
34 -1 V
34 0 V
34 -1 V
34 -1 V
34 0 V
34 -1 V
35 -1 V
35 0 V
34 -1 V
35 -1 V
35 0 V
35 -1 V
35 0 V
36 -1 V
35 -1 V
35 0 V
36 -1 V
36 0 V
35 -1 V
36 -1 V
36 0 V
36 -1 V
36 0 V
36 -1 V
37 0 V
36 -1 V
36 0 V
37 -1 V
36 0 V
37 -1 V
36 0 V
37 -1 V
37 0 V
36 -1 V
37 0 V
37 -1 V
37 0 V
37 -1 V
37 0 V
36 0 V
37 -1 V
37 0 V
37 -1 V
37 0 V
38 -1 V
37 0 V
37 0 V
37 -1 V
37 0 V
37 -1 V
37 0 V
37 0 V
34 -1 V
1.000 UP
1.000 UL
LT3
438 1850 M
0 58 V
-31 -58 R
62 0 V
-62 58 R
62 0 V
80 -792 R
0 30 V
-31 -30 R
62 0 V
-62 30 R
62 0 V
722 821 M
0 18 V
691 821 M
62 0 V
-62 18 R
62 0 V
947 678 M
0 18 V
916 678 M
62 0 V
-62 18 R
62 0 V
205 -95 R
0 20 V
-31 -20 R
62 0 V
-62 20 R
62 0 V
381 -76 R
0 15 V
-31 -15 R
62 0 V
-62 15 R
62 0 V
402 -50 R
0 11 V
-31 -11 R
62 0 V
-62 11 R
62 0 V
691 -53 R
0 11 V
-31 -11 R
62 0 V
-62 11 R
62 0 V
438 1879 Pls
549 1131 Pls
722 830 Pls
947 687 Pls
1183 611 Pls
1595 553 Pls
2028 515 Pls
2750 473 Pls
stroke
grestore
end
showpage
}}%
\put(1925,50){\makebox(0,0){$\frac{T}{T_c}$}}%
\put(100,1180){%
\makebox(0,0)[b]{\shortstack{$\frac{S_w}{L_xL_y}$}}%
}%
\put(3450,200){\makebox(0,0){2.6}}%
\put(3069,200){\makebox(0,0){2.4}}%
\put(2687,200){\makebox(0,0){2.2}}%
\put(2306,200){\makebox(0,0){2}}%
\put(1925,200){\makebox(0,0){1.8}}%
\put(1544,200){\makebox(0,0){1.6}}%
\put(1163,200){\makebox(0,0){1.4}}%
\put(781,200){\makebox(0,0){1.2}}%
\put(400,200){\makebox(0,0){1}}%
\put(350,2060){\makebox(0,0)[r]{1.8}}%
\put(350,1864){\makebox(0,0)[r]{1.6}}%
\put(350,1669){\makebox(0,0)[r]{1.4}}%
\put(350,1473){\makebox(0,0)[r]{1.2}}%
\put(350,1278){\makebox(0,0)[r]{1}}%
\put(350,1082){\makebox(0,0)[r]{0.8}}%
\put(350,887){\makebox(0,0)[r]{0.6}}%
\put(350,691){\makebox(0,0)[r]{0.4}}%
\put(350,496){\makebox(0,0)[r]{0.2}}%
\put(350,300){\makebox(0,0)[r]{0}}%
\end{picture}%
\endgroup
 
}
\end	{center}
\vspace{-7mm}
\caption{Action per unit area of the $k=1$ domain wall in SU(4)
with $aT=0.25$. Monte Carlo values, $+$, compared with 
perturbation theory based on various couplings:
$g^2(a)$, solid line, $g^2_I(a)$, long dashed line, $g^2_{SF}(T)$,
short  dashed line.}
\label{fig_wallPT}
\end 	{figure}

\begin	{figure}[p]
\begin	{center}
\leavevmode
\scalebox{0.8}{
\begingroup%
  \makeatletter%
  \newcommand{\GNUPLOTspecial}{%
    \@sanitize\catcode`\%=14\relax\special}%
  \setlength{\unitlength}{0.1bp}%
{\GNUPLOTspecial{!
/gnudict 256 dict def
gnudict begin
/Color false def
/Solid false def
/gnulinewidth 5.000 def
/userlinewidth gnulinewidth def
/vshift -33 def
/dl {10 mul} def
/hpt_ 31.5 def
/vpt_ 31.5 def
/hpt hpt_ def
/vpt vpt_ def
/M {moveto} bind def
/L {lineto} bind def
/R {rmoveto} bind def
/V {rlineto} bind def
/vpt2 vpt 2 mul def
/hpt2 hpt 2 mul def
/Lshow { currentpoint stroke M
  0 vshift R show } def
/Rshow { currentpoint stroke M
  dup stringwidth pop neg vshift R show } def
/Cshow { currentpoint stroke M
  dup stringwidth pop -2 div vshift R show } def
/UP { dup vpt_ mul /vpt exch def hpt_ mul /hpt exch def
  /hpt2 hpt 2 mul def /vpt2 vpt 2 mul def } def
/DL { Color {setrgbcolor Solid {pop []} if 0 setdash }
 {pop pop pop Solid {pop []} if 0 setdash} ifelse } def
/BL { stroke userlinewidth 2 mul setlinewidth } def
/AL { stroke userlinewidth 2 div setlinewidth } def
/UL { dup gnulinewidth mul /userlinewidth exch def
      10 mul /udl exch def } def
/PL { stroke userlinewidth setlinewidth } def
/LTb { BL [] 0 0 0 DL } def
/LTa { AL [1 udl mul 2 udl mul] 0 setdash 0 0 0 setrgbcolor } def
/LT0 { PL [] 1 0 0 DL } def
/LT1 { PL [4 dl 2 dl] 0 1 0 DL } def
/LT2 { PL [2 dl 3 dl] 0 0 1 DL } def
/LT3 { PL [1 dl 1.5 dl] 1 0 1 DL } def
/LT4 { PL [5 dl 2 dl 1 dl 2 dl] 0 1 1 DL } def
/LT5 { PL [4 dl 3 dl 1 dl 3 dl] 1 1 0 DL } def
/LT6 { PL [2 dl 2 dl 2 dl 4 dl] 0 0 0 DL } def
/LT7 { PL [2 dl 2 dl 2 dl 2 dl 2 dl 4 dl] 1 0.3 0 DL } def
/LT8 { PL [2 dl 2 dl 2 dl 2 dl 2 dl 2 dl 2 dl 4 dl] 0.5 0.5 0.5 DL } def
/Pnt { stroke [] 0 setdash
   gsave 1 setlinecap M 0 0 V stroke grestore } def
/Dia { stroke [] 0 setdash 2 copy vpt add M
  hpt neg vpt neg V hpt vpt neg V
  hpt vpt V hpt neg vpt V closepath stroke
  Pnt } def
/Pls { stroke [] 0 setdash vpt sub M 0 vpt2 V
  currentpoint stroke M
  hpt neg vpt neg R hpt2 0 V stroke
  } def
/Box { stroke [] 0 setdash 2 copy exch hpt sub exch vpt add M
  0 vpt2 neg V hpt2 0 V 0 vpt2 V
  hpt2 neg 0 V closepath stroke
  Pnt } def
/Crs { stroke [] 0 setdash exch hpt sub exch vpt add M
  hpt2 vpt2 neg V currentpoint stroke M
  hpt2 neg 0 R hpt2 vpt2 V stroke } def
/TriU { stroke [] 0 setdash 2 copy vpt 1.12 mul add M
  hpt neg vpt -1.62 mul V
  hpt 2 mul 0 V
  hpt neg vpt 1.62 mul V closepath stroke
  Pnt  } def
/Star { 2 copy Pls Crs } def
/BoxF { stroke [] 0 setdash exch hpt sub exch vpt add M
  0 vpt2 neg V  hpt2 0 V  0 vpt2 V
  hpt2 neg 0 V  closepath fill } def
/TriUF { stroke [] 0 setdash vpt 1.12 mul add M
  hpt neg vpt -1.62 mul V
  hpt 2 mul 0 V
  hpt neg vpt 1.62 mul V closepath fill } def
/TriD { stroke [] 0 setdash 2 copy vpt 1.12 mul sub M
  hpt neg vpt 1.62 mul V
  hpt 2 mul 0 V
  hpt neg vpt -1.62 mul V closepath stroke
  Pnt  } def
/TriDF { stroke [] 0 setdash vpt 1.12 mul sub M
  hpt neg vpt 1.62 mul V
  hpt 2 mul 0 V
  hpt neg vpt -1.62 mul V closepath fill} def
/DiaF { stroke [] 0 setdash vpt add M
  hpt neg vpt neg V hpt vpt neg V
  hpt vpt V hpt neg vpt V closepath fill } def
/Pent { stroke [] 0 setdash 2 copy gsave
  translate 0 hpt M 4 {72 rotate 0 hpt L} repeat
  closepath stroke grestore Pnt } def
/PentF { stroke [] 0 setdash gsave
  translate 0 hpt M 4 {72 rotate 0 hpt L} repeat
  closepath fill grestore } def
/Circle { stroke [] 0 setdash 2 copy
  hpt 0 360 arc stroke Pnt } def
/CircleF { stroke [] 0 setdash hpt 0 360 arc fill } def
/C0 { BL [] 0 setdash 2 copy moveto vpt 90 450  arc } bind def
/C1 { BL [] 0 setdash 2 copy        moveto
       2 copy  vpt 0 90 arc closepath fill
               vpt 0 360 arc closepath } bind def
/C2 { BL [] 0 setdash 2 copy moveto
       2 copy  vpt 90 180 arc closepath fill
               vpt 0 360 arc closepath } bind def
/C3 { BL [] 0 setdash 2 copy moveto
       2 copy  vpt 0 180 arc closepath fill
               vpt 0 360 arc closepath } bind def
/C4 { BL [] 0 setdash 2 copy moveto
       2 copy  vpt 180 270 arc closepath fill
               vpt 0 360 arc closepath } bind def
/C5 { BL [] 0 setdash 2 copy moveto
       2 copy  vpt 0 90 arc
       2 copy moveto
       2 copy  vpt 180 270 arc closepath fill
               vpt 0 360 arc } bind def
/C6 { BL [] 0 setdash 2 copy moveto
      2 copy  vpt 90 270 arc closepath fill
              vpt 0 360 arc closepath } bind def
/C7 { BL [] 0 setdash 2 copy moveto
      2 copy  vpt 0 270 arc closepath fill
              vpt 0 360 arc closepath } bind def
/C8 { BL [] 0 setdash 2 copy moveto
      2 copy vpt 270 360 arc closepath fill
              vpt 0 360 arc closepath } bind def
/C9 { BL [] 0 setdash 2 copy moveto
      2 copy  vpt 270 450 arc closepath fill
              vpt 0 360 arc closepath } bind def
/C10 { BL [] 0 setdash 2 copy 2 copy moveto vpt 270 360 arc closepath fill
       2 copy moveto
       2 copy vpt 90 180 arc closepath fill
               vpt 0 360 arc closepath } bind def
/C11 { BL [] 0 setdash 2 copy moveto
       2 copy  vpt 0 180 arc closepath fill
       2 copy moveto
       2 copy  vpt 270 360 arc closepath fill
               vpt 0 360 arc closepath } bind def
/C12 { BL [] 0 setdash 2 copy moveto
       2 copy  vpt 180 360 arc closepath fill
               vpt 0 360 arc closepath } bind def
/C13 { BL [] 0 setdash  2 copy moveto
       2 copy  vpt 0 90 arc closepath fill
       2 copy moveto
       2 copy  vpt 180 360 arc closepath fill
               vpt 0 360 arc closepath } bind def
/C14 { BL [] 0 setdash 2 copy moveto
       2 copy  vpt 90 360 arc closepath fill
               vpt 0 360 arc } bind def
/C15 { BL [] 0 setdash 2 copy vpt 0 360 arc closepath fill
               vpt 0 360 arc closepath } bind def
/Rec   { newpath 4 2 roll moveto 1 index 0 rlineto 0 exch rlineto
       neg 0 rlineto closepath } bind def
/Square { dup Rec } bind def
/Bsquare { vpt sub exch vpt sub exch vpt2 Square } bind def
/S0 { BL [] 0 setdash 2 copy moveto 0 vpt rlineto BL Bsquare } bind def
/S1 { BL [] 0 setdash 2 copy vpt Square fill Bsquare } bind def
/S2 { BL [] 0 setdash 2 copy exch vpt sub exch vpt Square fill Bsquare } bind def
/S3 { BL [] 0 setdash 2 copy exch vpt sub exch vpt2 vpt Rec fill Bsquare } bind def
/S4 { BL [] 0 setdash 2 copy exch vpt sub exch vpt sub vpt Square fill Bsquare } bind def
/S5 { BL [] 0 setdash 2 copy 2 copy vpt Square fill
       exch vpt sub exch vpt sub vpt Square fill Bsquare } bind def
/S6 { BL [] 0 setdash 2 copy exch vpt sub exch vpt sub vpt vpt2 Rec fill Bsquare } bind def
/S7 { BL [] 0 setdash 2 copy exch vpt sub exch vpt sub vpt vpt2 Rec fill
       2 copy vpt Square fill
       Bsquare } bind def
/S8 { BL [] 0 setdash 2 copy vpt sub vpt Square fill Bsquare } bind def
/S9 { BL [] 0 setdash 2 copy vpt sub vpt vpt2 Rec fill Bsquare } bind def
/S10 { BL [] 0 setdash 2 copy vpt sub vpt Square fill 2 copy exch vpt sub exch vpt Square fill
       Bsquare } bind def
/S11 { BL [] 0 setdash 2 copy vpt sub vpt Square fill 2 copy exch vpt sub exch vpt2 vpt Rec fill
       Bsquare } bind def
/S12 { BL [] 0 setdash 2 copy exch vpt sub exch vpt sub vpt2 vpt Rec fill Bsquare } bind def
/S13 { BL [] 0 setdash 2 copy exch vpt sub exch vpt sub vpt2 vpt Rec fill
       2 copy vpt Square fill Bsquare } bind def
/S14 { BL [] 0 setdash 2 copy exch vpt sub exch vpt sub vpt2 vpt Rec fill
       2 copy exch vpt sub exch vpt Square fill Bsquare } bind def
/S15 { BL [] 0 setdash 2 copy Bsquare fill Bsquare } bind def
/D0 { gsave translate 45 rotate 0 0 S0 stroke grestore } bind def
/D1 { gsave translate 45 rotate 0 0 S1 stroke grestore } bind def
/D2 { gsave translate 45 rotate 0 0 S2 stroke grestore } bind def
/D3 { gsave translate 45 rotate 0 0 S3 stroke grestore } bind def
/D4 { gsave translate 45 rotate 0 0 S4 stroke grestore } bind def
/D5 { gsave translate 45 rotate 0 0 S5 stroke grestore } bind def
/D6 { gsave translate 45 rotate 0 0 S6 stroke grestore } bind def
/D7 { gsave translate 45 rotate 0 0 S7 stroke grestore } bind def
/D8 { gsave translate 45 rotate 0 0 S8 stroke grestore } bind def
/D9 { gsave translate 45 rotate 0 0 S9 stroke grestore } bind def
/D10 { gsave translate 45 rotate 0 0 S10 stroke grestore } bind def
/D11 { gsave translate 45 rotate 0 0 S11 stroke grestore } bind def
/D12 { gsave translate 45 rotate 0 0 S12 stroke grestore } bind def
/D13 { gsave translate 45 rotate 0 0 S13 stroke grestore } bind def
/D14 { gsave translate 45 rotate 0 0 S14 stroke grestore } bind def
/D15 { gsave translate 45 rotate 0 0 S15 stroke grestore } bind def
/DiaE { stroke [] 0 setdash vpt add M
  hpt neg vpt neg V hpt vpt neg V
  hpt vpt V hpt neg vpt V closepath stroke } def
/BoxE { stroke [] 0 setdash exch hpt sub exch vpt add M
  0 vpt2 neg V hpt2 0 V 0 vpt2 V
  hpt2 neg 0 V closepath stroke } def
/TriUE { stroke [] 0 setdash vpt 1.12 mul add M
  hpt neg vpt -1.62 mul V
  hpt 2 mul 0 V
  hpt neg vpt 1.62 mul V closepath stroke } def
/TriDE { stroke [] 0 setdash vpt 1.12 mul sub M
  hpt neg vpt 1.62 mul V
  hpt 2 mul 0 V
  hpt neg vpt -1.62 mul V closepath stroke } def
/PentE { stroke [] 0 setdash gsave
  translate 0 hpt M 4 {72 rotate 0 hpt L} repeat
  closepath stroke grestore } def
/CircE { stroke [] 0 setdash 
  hpt 0 360 arc stroke } def
/Opaque { gsave closepath 1 setgray fill grestore 0 setgray closepath } def
/DiaW { stroke [] 0 setdash vpt add M
  hpt neg vpt neg V hpt vpt neg V
  hpt vpt V hpt neg vpt V Opaque stroke } def
/BoxW { stroke [] 0 setdash exch hpt sub exch vpt add M
  0 vpt2 neg V hpt2 0 V 0 vpt2 V
  hpt2 neg 0 V Opaque stroke } def
/TriUW { stroke [] 0 setdash vpt 1.12 mul add M
  hpt neg vpt -1.62 mul V
  hpt 2 mul 0 V
  hpt neg vpt 1.62 mul V Opaque stroke } def
/TriDW { stroke [] 0 setdash vpt 1.12 mul sub M
  hpt neg vpt 1.62 mul V
  hpt 2 mul 0 V
  hpt neg vpt -1.62 mul V Opaque stroke } def
/PentW { stroke [] 0 setdash gsave
  translate 0 hpt M 4 {72 rotate 0 hpt L} repeat
  Opaque stroke grestore } def
/CircW { stroke [] 0 setdash 
  hpt 0 360 arc Opaque stroke } def
/BoxFill { gsave Rec 1 setgray fill grestore } def
end
}}%
\begin{picture}(3600,2160)(0,0)%
{\GNUPLOTspecial{"
gnudict begin
gsave
0 0 translate
0.100 0.100 scale
0 setgray
newpath
1.000 UL
LTb
450 300 M
63 0 V
2937 0 R
-63 0 V
450 652 M
63 0 V
2937 0 R
-63 0 V
450 1004 M
63 0 V
2937 0 R
-63 0 V
450 1356 M
63 0 V
2937 0 R
-63 0 V
450 1708 M
63 0 V
2937 0 R
-63 0 V
450 2060 M
63 0 V
2937 0 R
-63 0 V
450 300 M
0 63 V
0 1697 R
0 -63 V
825 300 M
0 63 V
0 1697 R
0 -63 V
1200 300 M
0 63 V
0 1697 R
0 -63 V
1575 300 M
0 63 V
0 1697 R
0 -63 V
1950 300 M
0 63 V
0 1697 R
0 -63 V
2325 300 M
0 63 V
0 1697 R
0 -63 V
2700 300 M
0 63 V
0 1697 R
0 -63 V
3075 300 M
0 63 V
0 1697 R
0 -63 V
3450 300 M
0 63 V
0 1697 R
0 -63 V
1.000 UL
LTb
450 300 M
3000 0 V
0 1760 V
-3000 0 V
450 300 L
1.000 UP
1.000 UL
LT0
487 1011 M
0 366 V
456 1011 M
62 0 V
-62 366 R
62 0 V
78 -113 R
0 338 V
565 1264 M
62 0 V
-62 338 R
62 0 V
766 1405 M
0 324 V
735 1405 M
62 0 V
-62 324 R
62 0 V
988 1455 M
0 408 V
957 1455 M
62 0 V
-62 408 R
62 0 V
201 -507 R
0 535 V
-31 -535 R
62 0 V
-62 535 R
62 0 V
374 -739 R
0 479 V
-31 -479 R
62 0 V
-62 479 R
62 0 V
395 -500 R
0 380 V
-31 -380 R
62 0 V
-62 380 R
62 0 V
680 -56 R
0 549 V
-31 -549 R
62 0 V
-62 549 R
62 0 V
487 1194 Pls
596 1433 Pls
766 1567 Pls
988 1659 Pls
1220 1624 Pls
1625 1391 Pls
2051 1321 Pls
2762 1729 Pls
1.000 UL
LT1
450 1239 M
30 0 V
31 0 V
30 0 V
30 0 V
31 0 V
30 0 V
30 0 V
30 0 V
31 0 V
30 0 V
30 0 V
31 0 V
30 0 V
30 0 V
31 0 V
30 0 V
30 0 V
30 0 V
31 0 V
30 0 V
30 0 V
31 0 V
30 0 V
30 0 V
31 0 V
30 0 V
30 0 V
30 0 V
31 0 V
30 0 V
30 0 V
31 0 V
30 0 V
30 0 V
31 0 V
30 0 V
30 0 V
31 0 V
30 0 V
30 0 V
30 0 V
31 0 V
30 0 V
30 0 V
31 0 V
30 0 V
30 0 V
31 0 V
30 0 V
30 0 V
30 0 V
31 0 V
30 0 V
30 0 V
31 0 V
30 0 V
30 0 V
31 0 V
30 0 V
30 0 V
30 0 V
31 0 V
30 0 V
30 0 V
31 0 V
30 0 V
30 0 V
31 0 V
30 0 V
30 0 V
31 0 V
30 0 V
30 0 V
30 0 V
31 0 V
30 0 V
30 0 V
31 0 V
30 0 V
30 0 V
31 0 V
30 0 V
30 0 V
30 0 V
31 0 V
30 0 V
30 0 V
31 0 V
30 0 V
30 0 V
31 0 V
30 0 V
30 0 V
30 0 V
31 0 V
30 0 V
30 0 V
31 0 V
30 0 V
stroke
grestore
end
showpage
}}%
\put(1950,50){\makebox(0,0){$\frac{T}{T_c}$}}%
\put(100,1180){%
\makebox(0,0)[b]{\shortstack{$\frac{\Delta S_2}{\Delta S_1}$}}%
}%
\put(3450,200){\makebox(0,0){2.6}}%
\put(3075,200){\makebox(0,0){2.4}}%
\put(2700,200){\makebox(0,0){2.2}}%
\put(2325,200){\makebox(0,0){2}}%
\put(1950,200){\makebox(0,0){1.8}}%
\put(1575,200){\makebox(0,0){1.6}}%
\put(1200,200){\makebox(0,0){1.4}}%
\put(825,200){\makebox(0,0){1.2}}%
\put(450,200){\makebox(0,0){1}}%
\put(400,2060){\makebox(0,0)[r]{1.45}}%
\put(400,1708){\makebox(0,0)[r]{1.4}}%
\put(400,1356){\makebox(0,0)[r]{1.35}}%
\put(400,1004){\makebox(0,0)[r]{1.3}}%
\put(400,652){\makebox(0,0)[r]{1.25}}%
\put(400,300){\makebox(0,0)[r]{1.2}}%
\end{picture}%
\endgroup
 
}
\end	{center}
\vspace{-7mm}
\caption{Ratio $\Delta S^2_w/\Delta S^1_w$ in SU(4).}
\label{fig_ratios}
\end 	{figure}
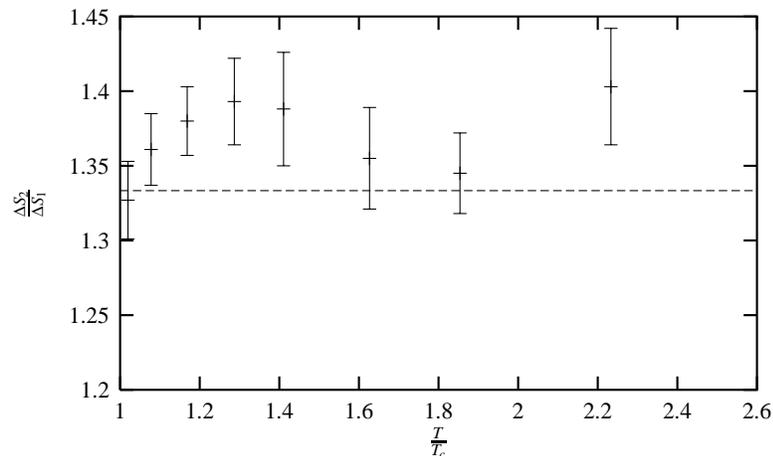

Our most extensive results are for SU(4) with $L_0=4$.
We show the values of $\Delta S^{k=1}_W$ for these calculations in
Fig.~\ref{fig_wallPT}, compared to the two-loop perturbative 
expectations using a mean-field
improved coupling and the 
bare coupling. We also include a `good'
coupling evaluated at low energy scales: the Schrodinger functional coupling 
which has been calculated non-perturbatively in 
\cite{alpha_s}.
We plot the ratios $\Delta S^2_w/\Delta S^1_w$ in
Fig.~\ref{fig_ratios}. 

We obtained results at one additional point,
not included in the graphs, at $\beta=20$, corresponding to $T \sim 1000T_c$.
Here we found excellent agreement with perturbation theory, as expected.

In Table~\ref{table_result} we list our results
for $\Delta S^k_W$ for SU(4) with $L_0=5$ and for SU(2), SU(3), and SU(6).
We include our results for SU(4) with $L_0=4$ at the 
same temperatures for comparison.

For SU(4) with $L_0=4$,
as $T\to T_c$ we find $\Delta S^k_W$ grows
much more rapidly than the perturbative prediction, reaching a factor of 18
at $T\simeq 1.02 T_c$. This tells us that
$\partial\sigma^k_W/\partial T$ becomes much larger than the 
low-order perturbative expectation as $T\to T_c$,
implying that $\sigma^k_W(T)$ is increasingly
suppressed relative to its perturbative value 
as we approach $T_c$. The ratio of the $\Delta S^k_W$
satisfies Casimir Scaling to good accuracy, though there is 
some evidence for a discrepancy below $1.5 T_c$.

To investigate the continuum limit we compare to $L_0=5$.
The discrepancy with perturbation theory is the same
at $T\simeq 1.88 T_c$ and similar at $T\simeq 1.02 T_c$.
It is clear that a large and
growing mismatch with perturbation theory as $T\to T_c$ is a 
feature of the continuum theory.
The ratio of the $\Delta S^k_W$ continues to
be close to Casimir Scaling, so that is 
also a property of the continuum theory.

For SU(6) we observe at both values of $T$
precisely the same discrepancy with perturbation theory as we 
saw for SU(4) at the same value of $L_0$. In addition the
$\Delta S^k_W$ ratios continue to satisfy Casimir Scaling.
Taken together this tells us that the derivative of the domain
wall tension has no factors of $k$ and $N$ except for
the Casimir scaling factor $k(N-k)$ and its dependence on
the 't Hooft coupling,  $g^2 N$.
Our SU(2) and SU(3) calculations
show a very similar discrepancy with perturbation theory.
Thus the suppression of  $\sigma^k_W(T)$
as $T\to T_c$ is largely independent of $N$.

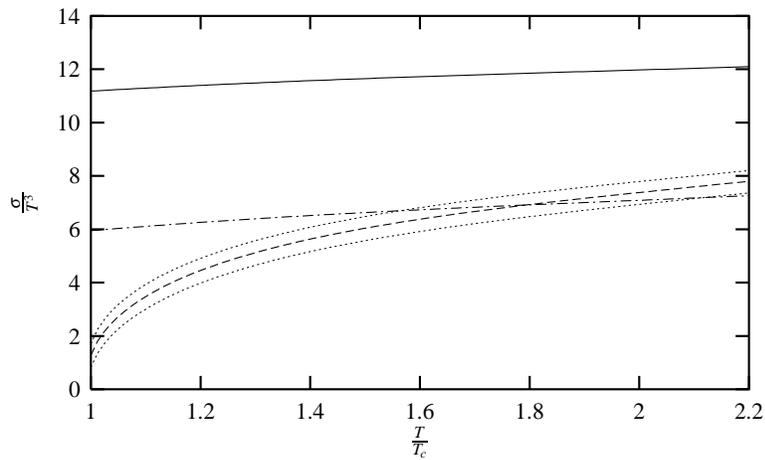
\begin	{figure}[p]
\begin	{center}
\leavevmode
\scalebox{0.8}{
\begingroup%
  \makeatletter%
  \newcommand{\GNUPLOTspecial}{%
    \@sanitize\catcode`\%=14\relax\special}%
  \setlength{\unitlength}{0.1bp}%
{\GNUPLOTspecial{!
/gnudict 256 dict def
gnudict begin
/Color false def
/Solid false def
/gnulinewidth 5.000 def
/userlinewidth gnulinewidth def
/vshift -33 def
/dl {10 mul} def
/hpt_ 31.5 def
/vpt_ 31.5 def
/hpt hpt_ def
/vpt vpt_ def
/M {moveto} bind def
/L {lineto} bind def
/R {rmoveto} bind def
/V {rlineto} bind def
/vpt2 vpt 2 mul def
/hpt2 hpt 2 mul def
/Lshow { currentpoint stroke M
  0 vshift R show } def
/Rshow { currentpoint stroke M
  dup stringwidth pop neg vshift R show } def
/Cshow { currentpoint stroke M
  dup stringwidth pop -2 div vshift R show } def
/UP { dup vpt_ mul /vpt exch def hpt_ mul /hpt exch def
  /hpt2 hpt 2 mul def /vpt2 vpt 2 mul def } def
/DL { Color {setrgbcolor Solid {pop []} if 0 setdash }
 {pop pop pop Solid {pop []} if 0 setdash} ifelse } def
/BL { stroke userlinewidth 2 mul setlinewidth } def
/AL { stroke userlinewidth 2 div setlinewidth } def
/UL { dup gnulinewidth mul /userlinewidth exch def
      10 mul /udl exch def } def
/PL { stroke userlinewidth setlinewidth } def
/LTb { BL [] 0 0 0 DL } def
/LTa { AL [1 udl mul 2 udl mul] 0 setdash 0 0 0 setrgbcolor } def
/LT0 { PL [] 1 0 0 DL } def
/LT1 { PL [4 dl 2 dl] 0 1 0 DL } def
/LT2 { PL [2 dl 3 dl] 0 0 1 DL } def
/LT3 { PL [1 dl 1.5 dl] 1 0 1 DL } def
/LT4 { PL [5 dl 2 dl 1 dl 2 dl] 0 1 1 DL } def
/LT5 { PL [4 dl 3 dl 1 dl 3 dl] 1 1 0 DL } def
/LT6 { PL [2 dl 2 dl 2 dl 4 dl] 0 0 0 DL } def
/LT7 { PL [2 dl 2 dl 2 dl 2 dl 2 dl 4 dl] 1 0.3 0 DL } def
/LT8 { PL [2 dl 2 dl 2 dl 2 dl 2 dl 2 dl 2 dl 4 dl] 0.5 0.5 0.5 DL } def
/Pnt { stroke [] 0 setdash
   gsave 1 setlinecap M 0 0 V stroke grestore } def
/Dia { stroke [] 0 setdash 2 copy vpt add M
  hpt neg vpt neg V hpt vpt neg V
  hpt vpt V hpt neg vpt V closepath stroke
  Pnt } def
/Pls { stroke [] 0 setdash vpt sub M 0 vpt2 V
  currentpoint stroke M
  hpt neg vpt neg R hpt2 0 V stroke
  } def
/Box { stroke [] 0 setdash 2 copy exch hpt sub exch vpt add M
  0 vpt2 neg V hpt2 0 V 0 vpt2 V
  hpt2 neg 0 V closepath stroke
  Pnt } def
/Crs { stroke [] 0 setdash exch hpt sub exch vpt add M
  hpt2 vpt2 neg V currentpoint stroke M
  hpt2 neg 0 R hpt2 vpt2 V stroke } def
/TriU { stroke [] 0 setdash 2 copy vpt 1.12 mul add M
  hpt neg vpt -1.62 mul V
  hpt 2 mul 0 V
  hpt neg vpt 1.62 mul V closepath stroke
  Pnt  } def
/Star { 2 copy Pls Crs } def
/BoxF { stroke [] 0 setdash exch hpt sub exch vpt add M
  0 vpt2 neg V  hpt2 0 V  0 vpt2 V
  hpt2 neg 0 V  closepath fill } def
/TriUF { stroke [] 0 setdash vpt 1.12 mul add M
  hpt neg vpt -1.62 mul V
  hpt 2 mul 0 V
  hpt neg vpt 1.62 mul V closepath fill } def
/TriD { stroke [] 0 setdash 2 copy vpt 1.12 mul sub M
  hpt neg vpt 1.62 mul V
  hpt 2 mul 0 V
  hpt neg vpt -1.62 mul V closepath stroke
  Pnt  } def
/TriDF { stroke [] 0 setdash vpt 1.12 mul sub M
  hpt neg vpt 1.62 mul V
  hpt 2 mul 0 V
  hpt neg vpt -1.62 mul V closepath fill} def
/DiaF { stroke [] 0 setdash vpt add M
  hpt neg vpt neg V hpt vpt neg V
  hpt vpt V hpt neg vpt V closepath fill } def
/Pent { stroke [] 0 setdash 2 copy gsave
  translate 0 hpt M 4 {72 rotate 0 hpt L} repeat
  closepath stroke grestore Pnt } def
/PentF { stroke [] 0 setdash gsave
  translate 0 hpt M 4 {72 rotate 0 hpt L} repeat
  closepath fill grestore } def
/Circle { stroke [] 0 setdash 2 copy
  hpt 0 360 arc stroke Pnt } def
/CircleF { stroke [] 0 setdash hpt 0 360 arc fill } def
/C0 { BL [] 0 setdash 2 copy moveto vpt 90 450  arc } bind def
/C1 { BL [] 0 setdash 2 copy        moveto
       2 copy  vpt 0 90 arc closepath fill
               vpt 0 360 arc closepath } bind def
/C2 { BL [] 0 setdash 2 copy moveto
       2 copy  vpt 90 180 arc closepath fill
               vpt 0 360 arc closepath } bind def
/C3 { BL [] 0 setdash 2 copy moveto
       2 copy  vpt 0 180 arc closepath fill
               vpt 0 360 arc closepath } bind def
/C4 { BL [] 0 setdash 2 copy moveto
       2 copy  vpt 180 270 arc closepath fill
               vpt 0 360 arc closepath } bind def
/C5 { BL [] 0 setdash 2 copy moveto
       2 copy  vpt 0 90 arc
       2 copy moveto
       2 copy  vpt 180 270 arc closepath fill
               vpt 0 360 arc } bind def
/C6 { BL [] 0 setdash 2 copy moveto
      2 copy  vpt 90 270 arc closepath fill
              vpt 0 360 arc closepath } bind def
/C7 { BL [] 0 setdash 2 copy moveto
      2 copy  vpt 0 270 arc closepath fill
              vpt 0 360 arc closepath } bind def
/C8 { BL [] 0 setdash 2 copy moveto
      2 copy vpt 270 360 arc closepath fill
              vpt 0 360 arc closepath } bind def
/C9 { BL [] 0 setdash 2 copy moveto
      2 copy  vpt 270 450 arc closepath fill
              vpt 0 360 arc closepath } bind def
/C10 { BL [] 0 setdash 2 copy 2 copy moveto vpt 270 360 arc closepath fill
       2 copy moveto
       2 copy vpt 90 180 arc closepath fill
               vpt 0 360 arc closepath } bind def
/C11 { BL [] 0 setdash 2 copy moveto
       2 copy  vpt 0 180 arc closepath fill
       2 copy moveto
       2 copy  vpt 270 360 arc closepath fill
               vpt 0 360 arc closepath } bind def
/C12 { BL [] 0 setdash 2 copy moveto
       2 copy  vpt 180 360 arc closepath fill
               vpt 0 360 arc closepath } bind def
/C13 { BL [] 0 setdash  2 copy moveto
       2 copy  vpt 0 90 arc closepath fill
       2 copy moveto
       2 copy  vpt 180 360 arc closepath fill
               vpt 0 360 arc closepath } bind def
/C14 { BL [] 0 setdash 2 copy moveto
       2 copy  vpt 90 360 arc closepath fill
               vpt 0 360 arc } bind def
/C15 { BL [] 0 setdash 2 copy vpt 0 360 arc closepath fill
               vpt 0 360 arc closepath } bind def
/Rec   { newpath 4 2 roll moveto 1 index 0 rlineto 0 exch rlineto
       neg 0 rlineto closepath } bind def
/Square { dup Rec } bind def
/Bsquare { vpt sub exch vpt sub exch vpt2 Square } bind def
/S0 { BL [] 0 setdash 2 copy moveto 0 vpt rlineto BL Bsquare } bind def
/S1 { BL [] 0 setdash 2 copy vpt Square fill Bsquare } bind def
/S2 { BL [] 0 setdash 2 copy exch vpt sub exch vpt Square fill Bsquare } bind def
/S3 { BL [] 0 setdash 2 copy exch vpt sub exch vpt2 vpt Rec fill Bsquare } bind def
/S4 { BL [] 0 setdash 2 copy exch vpt sub exch vpt sub vpt Square fill Bsquare } bind def
/S5 { BL [] 0 setdash 2 copy 2 copy vpt Square fill
       exch vpt sub exch vpt sub vpt Square fill Bsquare } bind def
/S6 { BL [] 0 setdash 2 copy exch vpt sub exch vpt sub vpt vpt2 Rec fill Bsquare } bind def
/S7 { BL [] 0 setdash 2 copy exch vpt sub exch vpt sub vpt vpt2 Rec fill
       2 copy vpt Square fill
       Bsquare } bind def
/S8 { BL [] 0 setdash 2 copy vpt sub vpt Square fill Bsquare } bind def
/S9 { BL [] 0 setdash 2 copy vpt sub vpt vpt2 Rec fill Bsquare } bind def
/S10 { BL [] 0 setdash 2 copy vpt sub vpt Square fill 2 copy exch vpt sub exch vpt Square fill
       Bsquare } bind def
/S11 { BL [] 0 setdash 2 copy vpt sub vpt Square fill 2 copy exch vpt sub exch vpt2 vpt Rec fill
       Bsquare } bind def
/S12 { BL [] 0 setdash 2 copy exch vpt sub exch vpt sub vpt2 vpt Rec fill Bsquare } bind def
/S13 { BL [] 0 setdash 2 copy exch vpt sub exch vpt sub vpt2 vpt Rec fill
       2 copy vpt Square fill Bsquare } bind def
/S14 { BL [] 0 setdash 2 copy exch vpt sub exch vpt sub vpt2 vpt Rec fill
       2 copy exch vpt sub exch vpt Square fill Bsquare } bind def
/S15 { BL [] 0 setdash 2 copy Bsquare fill Bsquare } bind def
/D0 { gsave translate 45 rotate 0 0 S0 stroke grestore } bind def
/D1 { gsave translate 45 rotate 0 0 S1 stroke grestore } bind def
/D2 { gsave translate 45 rotate 0 0 S2 stroke grestore } bind def
/D3 { gsave translate 45 rotate 0 0 S3 stroke grestore } bind def
/D4 { gsave translate 45 rotate 0 0 S4 stroke grestore } bind def
/D5 { gsave translate 45 rotate 0 0 S5 stroke grestore } bind def
/D6 { gsave translate 45 rotate 0 0 S6 stroke grestore } bind def
/D7 { gsave translate 45 rotate 0 0 S7 stroke grestore } bind def
/D8 { gsave translate 45 rotate 0 0 S8 stroke grestore } bind def
/D9 { gsave translate 45 rotate 0 0 S9 stroke grestore } bind def
/D10 { gsave translate 45 rotate 0 0 S10 stroke grestore } bind def
/D11 { gsave translate 45 rotate 0 0 S11 stroke grestore } bind def
/D12 { gsave translate 45 rotate 0 0 S12 stroke grestore } bind def
/D13 { gsave translate 45 rotate 0 0 S13 stroke grestore } bind def
/D14 { gsave translate 45 rotate 0 0 S14 stroke grestore } bind def
/D15 { gsave translate 45 rotate 0 0 S15 stroke grestore } bind def
/DiaE { stroke [] 0 setdash vpt add M
  hpt neg vpt neg V hpt vpt neg V
  hpt vpt V hpt neg vpt V closepath stroke } def
/BoxE { stroke [] 0 setdash exch hpt sub exch vpt add M
  0 vpt2 neg V hpt2 0 V 0 vpt2 V
  hpt2 neg 0 V closepath stroke } def
/TriUE { stroke [] 0 setdash vpt 1.12 mul add M
  hpt neg vpt -1.62 mul V
  hpt 2 mul 0 V
  hpt neg vpt 1.62 mul V closepath stroke } def
/TriDE { stroke [] 0 setdash vpt 1.12 mul sub M
  hpt neg vpt 1.62 mul V
  hpt 2 mul 0 V
  hpt neg vpt -1.62 mul V closepath stroke } def
/PentE { stroke [] 0 setdash gsave
  translate 0 hpt M 4 {72 rotate 0 hpt L} repeat
  closepath stroke grestore } def
/CircE { stroke [] 0 setdash 
  hpt 0 360 arc stroke } def
/Opaque { gsave closepath 1 setgray fill grestore 0 setgray closepath } def
/DiaW { stroke [] 0 setdash vpt add M
  hpt neg vpt neg V hpt vpt neg V
  hpt vpt V hpt neg vpt V Opaque stroke } def
/BoxW { stroke [] 0 setdash exch hpt sub exch vpt add M
  0 vpt2 neg V hpt2 0 V 0 vpt2 V
  hpt2 neg 0 V Opaque stroke } def
/TriUW { stroke [] 0 setdash vpt 1.12 mul add M
  hpt neg vpt -1.62 mul V
  hpt 2 mul 0 V
  hpt neg vpt 1.62 mul V Opaque stroke } def
/TriDW { stroke [] 0 setdash vpt 1.12 mul sub M
  hpt neg vpt 1.62 mul V
  hpt 2 mul 0 V
  hpt neg vpt -1.62 mul V Opaque stroke } def
/PentW { stroke [] 0 setdash gsave
  translate 0 hpt M 4 {72 rotate 0 hpt L} repeat
  Opaque stroke grestore } def
/CircW { stroke [] 0 setdash 
  hpt 0 360 arc Opaque stroke } def
/BoxFill { gsave Rec 1 setgray fill grestore } def
end
}}%
\begin{picture}(3600,2160)(0,0)%
{\GNUPLOTspecial{"
gnudict begin
gsave
0 0 translate
0.100 0.100 scale
0 setgray
newpath
1.000 UL
LTb
350 300 M
63 0 V
3037 0 R
-63 0 V
350 551 M
63 0 V
3037 0 R
-63 0 V
350 803 M
63 0 V
3037 0 R
-63 0 V
350 1054 M
63 0 V
3037 0 R
-63 0 V
350 1306 M
63 0 V
3037 0 R
-63 0 V
350 1557 M
63 0 V
3037 0 R
-63 0 V
350 1809 M
63 0 V
3037 0 R
-63 0 V
350 2060 M
63 0 V
3037 0 R
-63 0 V
350 300 M
0 63 V
0 1697 R
0 -63 V
867 300 M
0 63 V
0 1697 R
0 -63 V
1383 300 M
0 63 V
0 1697 R
0 -63 V
1900 300 M
0 63 V
0 1697 R
0 -63 V
2417 300 M
0 63 V
0 1697 R
0 -63 V
2933 300 M
0 63 V
0 1697 R
0 -63 V
3450 300 M
0 63 V
0 1697 R
0 -63 V
1.000 UL
LTb
350 300 M
3100 0 V
0 1760 V
-3100 0 V
350 300 L
1.000 UL
LT0
350 1706 M
31 1 V
32 2 V
31 2 V
31 2 V
32 1 V
31 2 V
31 2 V
32 1 V
31 2 V
31 1 V
31 2 V
32 2 V
31 1 V
31 2 V
32 1 V
31 2 V
31 1 V
32 2 V
31 1 V
31 2 V
32 1 V
31 1 V
31 2 V
32 1 V
31 1 V
31 2 V
31 1 V
32 1 V
31 2 V
31 1 V
32 1 V
31 2 V
31 1 V
32 1 V
31 1 V
31 1 V
32 2 V
31 1 V
31 1 V
32 1 V
31 1 V
31 1 V
31 2 V
32 1 V
31 1 V
31 1 V
32 1 V
31 1 V
31 1 V
32 1 V
31 1 V
31 1 V
32 1 V
31 1 V
31 1 V
32 1 V
31 1 V
31 1 V
31 1 V
32 1 V
31 1 V
31 1 V
32 1 V
31 1 V
31 1 V
32 1 V
31 1 V
31 1 V
32 1 V
31 1 V
31 1 V
32 1 V
31 1 V
31 0 V
31 1 V
32 1 V
31 1 V
31 1 V
32 1 V
31 1 V
31 1 V
32 1 V
31 1 V
31 1 V
32 1 V
31 0 V
31 1 V
32 1 V
31 1 V
31 1 V
31 1 V
32 1 V
31 1 V
31 1 V
32 1 V
31 1 V
31 1 V
32 1 V
31 1 V
1.000 UL
LT1
350 460 M
31 66 V
32 46 V
31 36 V
31 31 V
32 27 V
31 24 V
31 22 V
32 20 V
31 19 V
31 17 V
31 17 V
32 15 V
31 15 V
31 13 V
32 14 V
31 12 V
31 12 V
32 11 V
31 11 V
31 11 V
32 10 V
31 10 V
31 9 V
32 9 V
31 9 V
31 9 V
31 8 V
32 8 V
31 8 V
31 7 V
32 8 V
31 7 V
31 7 V
32 7 V
31 6 V
31 7 V
32 6 V
31 6 V
31 6 V
32 6 V
31 6 V
31 5 V
31 6 V
32 5 V
31 5 V
31 5 V
32 6 V
31 4 V
31 5 V
32 5 V
31 5 V
31 4 V
32 5 V
31 4 V
31 5 V
32 4 V
31 4 V
31 4 V
31 4 V
32 4 V
31 4 V
31 4 V
32 4 V
31 4 V
31 4 V
32 4 V
31 3 V
31 4 V
32 4 V
31 3 V
31 4 V
32 3 V
31 4 V
31 3 V
31 4 V
32 3 V
31 3 V
31 4 V
32 3 V
31 3 V
31 4 V
32 3 V
31 3 V
31 3 V
32 4 V
31 3 V
31 3 V
32 3 V
31 4 V
31 3 V
31 3 V
32 3 V
31 3 V
31 4 V
32 3 V
31 3 V
31 3 V
32 3 V
31 4 V
0.500 UL
LT3
350 514 M
31 68 V
32 46 V
31 37 V
31 31 V
32 27 V
31 24 V
31 22 V
32 21 V
31 18 V
31 18 V
31 16 V
32 15 V
31 15 V
31 14 V
32 13 V
31 12 V
31 12 V
32 11 V
31 11 V
31 11 V
32 10 V
31 9 V
31 10 V
32 9 V
31 9 V
31 8 V
31 8 V
32 8 V
31 8 V
31 7 V
32 8 V
31 7 V
31 7 V
32 6 V
31 7 V
31 6 V
32 6 V
31 6 V
31 6 V
32 6 V
31 6 V
31 5 V
31 6 V
32 5 V
31 5 V
31 5 V
32 5 V
31 5 V
31 5 V
32 4 V
31 5 V
31 4 V
32 5 V
31 4 V
31 5 V
32 4 V
31 4 V
31 4 V
31 4 V
32 4 V
31 4 V
31 4 V
32 4 V
31 3 V
31 4 V
32 4 V
31 3 V
31 4 V
32 3 V
31 4 V
31 3 V
32 4 V
31 3 V
31 4 V
31 3 V
32 3 V
31 4 V
31 3 V
32 3 V
31 3 V
31 4 V
32 3 V
31 3 V
31 3 V
32 3 V
31 3 V
31 4 V
32 3 V
31 3 V
31 3 V
31 3 V
32 3 V
31 3 V
31 3 V
32 4 V
31 3 V
31 3 V
32 3 V
31 3 V
0.500 UL
LT3
350 404 M
31 65 V
32 45 V
31 36 V
31 31 V
32 27 V
31 23 V
31 22 V
32 20 V
31 19 V
31 17 V
31 16 V
32 16 V
31 14 V
31 14 V
32 13 V
31 12 V
31 12 V
32 12 V
31 11 V
31 10 V
32 10 V
31 10 V
31 10 V
32 9 V
31 9 V
31 8 V
31 9 V
32 8 V
31 8 V
31 7 V
32 8 V
31 7 V
31 7 V
32 7 V
31 6 V
31 7 V
32 6 V
31 6 V
31 6 V
32 6 V
31 6 V
31 6 V
31 5 V
32 6 V
31 5 V
31 5 V
32 5 V
31 5 V
31 5 V
32 5 V
31 5 V
31 5 V
32 4 V
31 5 V
31 4 V
32 4 V
31 5 V
31 4 V
31 4 V
32 4 V
31 4 V
31 4 V
32 4 V
31 4 V
31 4 V
32 4 V
31 3 V
31 4 V
32 4 V
31 3 V
31 4 V
32 4 V
31 3 V
31 4 V
31 3 V
32 4 V
31 3 V
31 3 V
32 4 V
31 3 V
31 4 V
32 3 V
31 3 V
31 4 V
32 3 V
31 3 V
31 3 V
32 4 V
31 3 V
31 3 V
31 4 V
32 3 V
31 3 V
31 3 V
32 4 V
31 3 V
31 3 V
32 3 V
31 4 V
1.000 UL
LT4
350 1048 M
31 2 V
32 3 V
31 2 V
31 3 V
32 2 V
31 3 V
31 2 V
32 2 V
31 3 V
31 2 V
31 2 V
32 3 V
31 2 V
31 2 V
32 2 V
31 3 V
31 2 V
32 2 V
31 2 V
31 2 V
32 2 V
31 2 V
31 2 V
32 2 V
31 2 V
31 2 V
31 2 V
32 2 V
31 2 V
31 2 V
32 1 V
31 2 V
31 2 V
32 2 V
31 2 V
31 1 V
32 2 V
31 2 V
31 1 V
32 2 V
31 2 V
31 1 V
31 2 V
32 2 V
31 1 V
31 2 V
32 1 V
31 2 V
31 1 V
32 2 V
31 1 V
31 2 V
32 1 V
31 2 V
31 1 V
32 2 V
31 1 V
31 1 V
31 2 V
32 1 V
31 2 V
31 1 V
32 1 V
31 2 V
31 1 V
32 1 V
31 2 V
31 1 V
32 1 V
31 2 V
31 1 V
32 1 V
31 1 V
31 2 V
31 1 V
32 1 V
31 2 V
31 1 V
32 1 V
31 2 V
31 1 V
32 1 V
31 1 V
31 2 V
32 1 V
31 1 V
31 2 V
32 1 V
31 1 V
31 2 V
31 1 V
32 1 V
31 2 V
31 1 V
32 1 V
31 2 V
31 1 V
32 1 V
31 2 V
stroke
grestore
end
showpage
}}%
\put(1900,50){\makebox(0,0){$\frac{T}{T_c}$}}%
\put(100,1180){%
\makebox(0,0)[b]{\shortstack{$\frac{\sigma}{T^3}$}}%
}%
\put(3450,200){\makebox(0,0){2.2}}%
\put(2933,200){\makebox(0,0){2}}%
\put(2417,200){\makebox(0,0){1.8}}%
\put(1900,200){\makebox(0,0){1.6}}%
\put(1383,200){\makebox(0,0){1.4}}%
\put(867,200){\makebox(0,0){1.2}}%
\put(350,200){\makebox(0,0){1}}%
\put(300,2060){\makebox(0,0)[r]{14}}%
\put(300,1809){\makebox(0,0)[r]{12}}%
\put(300,1557){\makebox(0,0)[r]{10}}%
\put(300,1306){\makebox(0,0)[r]{8}}%
\put(300,1054){\makebox(0,0)[r]{6}}%
\put(300,803){\makebox(0,0)[r]{4}}%
\put(300,551){\makebox(0,0)[r]{2}}%
\put(300,300){\makebox(0,0)[r]{0}}%
\end{picture}%
\endgroup
 
}
\end	{center}
\vspace{-7mm}
\caption{Surface tension in units of $T$, using the functional form in 
eqn(3.1)
fitted to our results.  For comparison we show the 1-loop (solid line) and 2-loop (dot-dashed line) perturbative results using a mean field improved coupling. All for the $k=1$ wall in SU(4).}
\label{fig_wallT}
\end 	{figure}

To estimate the suppression of $\sigma^k_W$ we can 
in principle calculate it
by interpolating $\partial \sigma^k_W/\partial\beta$
in $\beta$ and then integrating from large $\beta$
 down to the desired
value of $T$, but our calculations are 
not dense enough in $\beta$ for this.
We can obtain a qualitative picture by assuming
some functional form for $\partial \sigma^k_W/\partial\beta$ and 
fitting it to the calculated values. 
Choosing the $L_0=4$ SU(4) calculation, we make a fit
using a simple modification of the one-loop formula for the
$k$-wall free energy: 
\begin{equation}
\frac{F^k_W}{L_1L_2T}
=
[\mathrm{1 \ loop}]
+
a\exp(-b\sqrt{\beta_I-\beta_{Ic}})
\label{eqn_Ffit}
\end{equation}
Using this fit we obtain the $T$-dependence of $\sigma^k_W$ 
shown in Fig.~\ref{fig_wallT}. We display the uncertainty
from the errors in the fitted parameters, 
but it is clear that we cannot reliably estimate the systematic 
error inherent in the choice of fitting function.
The qualitative picture is that $\sigma^k_W$ is strongly
suppressed at $T\leq 1.5T_c$.
This suggests that it may reach zero at some $T_{\tilde{H}}$
somewhat below $T_c$, causing a second-order 
phase transition due to the condensation of spatial 't\nolinebreak Hooft 
strings.


\subsection{Profile}
We can average the Polyakov loop over the transverse coordinates to obtain
the profile $\bar{l}_p(x_3)$. Comparing with the perturbative prediction
\cite{CKA1,DWd3} we find good agreement down to $T\simeq 1.88 T_c$.
However, at $T\simeq 1.02 T_c$ the profile is very far from
the perturbative expectation. This observation
makes it all the more remarkable that we continue to see
Casimir Scaling at such very low $T$.

\subsection{Wetting}
A $k$-wall can interpolate between two deconfined phases by passing
through the origin of the complex $l_p(\vec{n})$ plane, 
i.e. the confined phase.
Whether this will happen or not depends on the relevant surface tensions.
We investigated this by a series of runs very near $T_c$ in SU(4) and SU(6).
We found that all the walls split into a pair of confined-deconfined walls
('wetting') over a range of $\sim 0.01 T_c$ in $T$. This small range
occurs because the latent heat is much larger than the domain wall tensions,
which suppresses the breaking up unless $T$ is extremely close to $T_c$.

\section{Conclusions}
We have shown that surface tensions of domain walls seperating
different deconfined phases are close to Casimir Scaling, i.e.
\begin{equation}
\sigma^k_W
=
k(N-k)\, f(g^2(T)N,T/T_c)\, T^3 ,
\label{eqn_factor1}
\end{equation}
to a good approximation.
The surface tensions are strongly suppressed as $T\to T_c$.
The domain wall profile is far from the perturbative prediction near $T_c$.
We also observed 'wetting' of the domain walls very near to $T_c$.



\begin{table}
\begin{center}
\footnotesize{
\begin{tabular}{|c|c|c|c|c|c|c|c|} \hline
N & $aT$ & k & Prediction & $\Delta S^k_w/L_xL_y$ & 
$\Delta S^k_w/\Delta S^1_w$ & CS & $T/T_c$ \\ \hline
2 & 0.25 & 1 & 0.1042 & 0.3089(84) & & & 1.88 \\ \hline
3 & 0.25 & 1 & 0.1032 & 1.380(40)  & & & 1.02 \\ 
3 & 0.25 & 1 & 0.0936 & 0.2716(67) & & & 1.88 \\ \hline
4 & 0.25 & 1 & 0.0867 & 1.615(30)  & & & 1.02 \\
  &      & 2 & 0.1157 & 2.141(30)  & 1.326(29) & 1.333 & 1.02 \\
4 & 0.25 & 1 & 0.0791 & 0.2200(56) & & & 1.88 \\
  &      & 2 & 0.1055 & 0.2957(57) & 1.344(31) & 1.333 & 1.88 \\ 
4 & 0.20 & 1 & 0.0514 & 0.739(27)  & & & 1.02 \\
  &      & 2 & 0.0685 & 0.982(28)  & 1.329(56) & 1.333 & 1.02 \\
4 & 0.20 & 1 & 0.0467 & 0.1292(54) & & & 1.88 \\
  &      & 2 & 0.0622 & 0.1743(60) & 1.350(57) & 1.333 & 1.88 \\ \hline
6 & 0.20 & 1 & 0.0381 & 0.488(25)  & & & 1.02 \\
  &      & 2 & 0.0610 & 0.847(27)  & 1.74(9)   & 1.60 & 1.02 \\
  &      & 3 & 0.0687 & 1.015(29)  & 2.08(10)  & 1.80 & 1.02 \\ 
6 & 0.20 & 1 & 0.0345 & 0.098(9)   & & & 1.88 \\
  &      & 2  & 0.0552 & 0.159(10)  & 1.62(14) & 1.60 & 1.88 \\
  &      & 3  & 0.0621 & 0.171(10)  & 1.74(15) & 1.80 & 1.88 \\ \hline
\end{tabular}
}
\caption{Results for $\Delta S^k_w$. The prediction is the two-loop perturbative expectation, using a mean-field improved coupling.}
\label{table_result}
\end{center}
\end{table}

\end{document}